\documentclass{aa}
\usepackage[dvips]{graphicx}
\def\simless{\mathbin{\lower 3pt\hbox
{$\rlap{\raise 5pt\hbox{$\char'074$}}\mathchar"7218$}}}   
\def\simmore{\mathbin{\lower 3pt\hbox
{$\rlap{\raise 5pt\hbox{$\char'076$}}\mathchar"7218$}}}   
\newcommand{\be}{\begin{equation}}
\newcommand{\ee}{\end{equation}}

\begin{document}
\title{Prompt GRB emission from gradual energy dissipation}
\titlerunning{Prompt GRB emission from gradual energy dissipation}
\authorrunning{Dimitrios Giannios}
\author{Dimitrios Giannios}

\institute{Max Planck Institute for Astrophysics, Box 1317, D-85741 Garching, Germany}

\offprints{giannios@mpa-garching.mpg.de}
\date{Received / Accepted}

\abstract
{I calculate the emission expected from a Poynting-flux-dominated
gamma-ray burst (GRB) flow in which energy is dissipated gradually by 
magnetic reconnection. In this picture, the energy of the radiating 
particles is determined by heating and cooling balance (slow heating model). 
Detailed radiative transfer calculations show that,
at Thomson optical depths of order of unity, the dominant radiative 
process is inverse Compton scattering. Synchrotron-self-absorbed emission and inverse 
Compton dominate in the Thomson thin parts of the flow. 
The electrons stay thermal throughout the dissipation region 
because of Coulomb collisions (Thomson thick part of the flow) 
and exchange of synchrotron photons (Thomson thin part). 
The resulting spectrum naturally explains the observed sub-MeV 
break of the GRB emission and the spectral slopes above and below the break. 
The model predicts that the  $\gamma$-ray power-law tail has a high-energy 
cutoff typically in the $\sim 0.1-1$ GeV energy range that should be
observable with {\it GLAST}. The model also predicts a prompt emission
component in the  optical and UV associated with the GeV emission. 
Observations of the prompt emission of GRB 061121 
that cover the energy range from the optical to $\sim 1$ MeV are explained by 
the model.    
\\   

}

\maketitle

\section{Introduction} 
\label{intro}

The GRB emission is the likely result of internal energy release
in an ultrarelativistic flow. The dissipative
and radiative mechanisms for the GRB largely remain uncertain.
A popular model for the energy dissipation invokes internal shocks 
in an unsteady flow (Paczynski \& Xu 1994; Rees \& M\'esz\'aros 1994).
An alternative proposal is magnetic dissipation in a 
strongly magnetized flow (Thompson 1994; Spruit et al. 2001)

Internal shocks are efficient in dissipating a large fraction
of the kinetic energy of the flow provided that it is highly variable,
i.e., composed of distinct ejection events with strong variation in their
bulk Lorentz factor $\gamma$ (e.g. Kobayashi et al. 1997). 
Energy is dissipated by the shocks at the location
of the collision of the shells. Particles are assumed to be 
accelerated on a very short timescale at the shock front to 
ultrarelativistic speeds and non-thermal distributions. 
Subsequently, they radiate a fraction of the dissipated energy
via synchrotron and inverse Compton processes. 
The relevant radiative mechanisms and the emitted spectra depend, to a large
extent, on the shock microphysics and the corresponding Thomson optical depth 
of the flow at the radius of the collision (see, e.g., Pe'er et al. 2006) 

On the other hand, the energy dissipation that powers the GRB emission 
may be gradual and distributed over a large part of the volume of the flow. 
The energy of the radiating particles is determined by the heating/cooling 
equilibrium (slow heating model; Ghisellini \& Celotti 1999; Stern \& Poutanen
2004). Such an energy balance is expected to lead to sub-relativistic or
mildly relativistic temperatures in the flow. 
Magnetic dissipation in a strongly magnetized flow can provide a
possible physical setup where gradual dissipation is realized.
As shown in Drenkhahn (2002) and Drenkhahn \& Spruit (2002; hereafter DS02), 
dissipation though reconnection takes place over several decades in radius; 
typically in both Thomson thick and thin conditions.
Dissipation in the reconnection model is responsible for both the 
acceleration of the flow and the prompt emission.

\subsection{Emission from magnetic dissipation}

In previous works (Giannios 2006; Giannios \& 
Spruit 2007; hereafter G06 and GS07 respectively), we studied the 
radiative transfer close to the Thomson photosphere of the flow in the context 
of the magnetic reconnection model. The detailed Monte Carlo calculations 
have shown that, due to energy release, the flow develops a hot photosphere 
with comoving electron temperatures of tens of keV. In the photospheric region, 
Coulomb collisions are fast enough to thermalize the electron distribution.
Upscattering of photons that are produced deeper in the flow
by those hot electrons leads to a powerful photospheric emission;
it accounts for $\sim3-30$\% of the luminosity of the flow.
The resulting  $E\cdot f(E)$ spectrum has a characteristic 
$\sim 1$ MeV peak followed by a flat high-energy power-law emission.      

Here, I extend the radiative transfer calculation to the Thomson 
thin region of the flow where, in the reconnection model,
there can be substantial energy release and associated emission.
Because of the strong magnetic fields synchrotron self absorption 
results in efficient energy exchange of the electrons, 
keeping their distribution thermal (Ghisellini et al. 1998). 
I show that heating and cooling balance in that region leads to electron 
temperatures of the order of the electron rest mass. Under these conditions, 
synchrotron-self-absorbed (SSA) emission is an important radiative mechanism
in addition to inverse Compton. SSA dominates the observed emission in the 
soft X-rays and softer bands. 

The efficient thermalization of the emitting particles throughout
the flow reduces the dependence of the model on the, poorly 
understood, mechanisms of particle acceleration that operate in 
magnetic reconnection. The model is defined by just 3 main 
parameters (luminosity, baryon loading and a reconnection-rate
parameter; see Sect. 3). In contrast to the internal shock model,
no quantities have to added to parameterize the particle distributions
and the amplification of magnetic fields; the field strength is an 
integral part of the reconnection model. Because the model is so well defined,
it makes direct and stable predictions for the emitted spectrum.
In the following, I show how this spectrum can be computed.

The structure of the paper is the following. In the next section
I summarize and contrast the main aspects of internal shock and magnetic reconnection
models for the prompt GRB emission. The dynamics of the flow in the reconnection model
and the radiative transfer in the flow are the topic of Sects. 3 and 4
respectively. The resulting spectra and their direct comparison with
multi-frequency observations (from optical
to $\gamma$-rays; see Page et al. 2007) of GRB 061121 are presented in
Sect. 5. I discuss the results and conclude in Sect. 6.

\section{Magnetic reconnection versus internal shocks}
\label{model}

The internal shock model for the GRB emission invokes high variability in the 
bulk Lorentz factor of the flow that leads to internal collisions.
The location where the collision of two shells takes place depends on their
initial separation and bulk Lorentz factors. 
These collisions can dissipate a substantial fraction of the kinetic 
energy of the flow. Internal shocks are assumed to lead to 
particle acceleration and magnetic field amplification
at the shock front. If electrons
receive a large fraction of the dissipated
energy then they are accelerated to ultrarelativistic speeds.
They cool down radiatively by synchrotron and inverse Compton
mechanisms. The resulting spectrum and the relevant radiative
mechanisms depend on details of the distribution of the accelerated
particles, the magnetic field strength and the Thomson optical depth of the 
flow at the location of the collision.
If the collision of two shells takes place at the Thomson thin region, as needed
to explain the typical variability properties of the GRB lightcurves
(Daigne \& Mochkovitch 1998; Nakar \& Piran 2002; Mimica et al. 2005), 
synchrotron self Compton is likely the most promising radiative mechanism
(Rees \& M\'esz\'aros 1994; Katz 1994; Tavani 1996). 
In this picture optically thin synchrotron emission dominates the observed 
hard X-ray, $\sim$MeV spectrum. Despite its attractive features,
the synchrotron model has theoretical and observational difficulties
(as discussed, for example, in  Ghirlanda et al. 2003). 
If, on the other hand, the collisions take place close to the Thomson
photosphere, Compton scattering is the dominant radiative mechanism that
shapes the spectrum and results in very different emission (Pe'er et al. 2006).

As an alternative to internal shocks, magnetic dissipation can power 
the prompt emission provided that the flow is launched Poynting-flux 
dominated (or with a substantial fraction of its energy
in the form of Poynting flux).
Magnetic dissipation through, for example, reconnection can release
energy smoothly in a large fraction of the volume of the flow. 
This energy release can take place while the flow expands over several decades in
radius (DS02). The energy of the radiating particles is determined by
balancing heating and radiative (or adiabatic) cooling
at each radius. In this case the slow heating picture described 
by Ghisellini \& Celotti (1999) takes place. The electrons
are subrelativistic or mildly relativistic and their synchrotron
emission is self absorbed. In a strongly magnetized flow, such SSA
emission guarantees efficient energy exchange and thermalization
of the electrons on a very short timescale (Ghisellini et al. 1998).
The resulting emission does not depend on details of particle acceleration
and magnetic field amplification that one faces in internal shock models.
The model is defined by just the luminosity, the baryon loading and the 
reconnection-rate parameter of the flow. This contributes significantly to 
the predictive power of the magnetic reconnection model. 
The total observed flux is the integrated emission 
from the different parts of the flow
in which dissipation of energy takes place. It contains both 
photospheric (Thompson 1994; Stern 1999; G06; GS07) and Thomson thin
components (this work).

\subsection{Implications from the observed GRB variablity}

One additional difference of the internal shock and the magnetic reconnection 
model is connected to implications from the observed variability of the lightcurves.
Internal shocks are efficient only in variable flows.
Variability and dissipation are, a priori, unrelated in the magnetic
reconnection model in which dissipation takes place, even in a steady 
outflow. On the other hand, the observed lightcurves are often highly variable
showing that the flow does evidently evolve during a GRB.
In the context of the reconnection model, the observed variability reflects changes in the
luminosity and baryon loading of the flow during the burst. 
As shown in GS07, the flow can be treated as
quasi-stationary for all but the shortest time scales observed in
a burst, the variation of spectral properties during a burst
directly reflects variations in the central engine. This is in
contrast to models in which the prompt radiation is produced at
much larger distances from the source, such as external shock
models. It is also in contrast with the internal shock model,
since the internal evolution of the flow between the source and
the level where radiation is emitted is a key ingredient in this
model. Deducing properties of the central engine from observed
burst properties is thus a much more direct prospect in the
magnetic dissipation model.

\section{Gradual energy release because of magnetic reconnection}

Magnetic dissipation can take place gradually in the GRB flow.
The rate at which energy is dissipated as a function of radius 
depends on the magnetic field geometry and the exact mechanism through
which magnetic energy dissipates. 
If the flow is launched with field of large scale, energy dissipation
can be a result of global MHD instabilities (such as current-driven
instabilities; e.g., Lyutikov \& Blandford 2003; Giannios \& Spruit 2006). 
On the other hand, if the flow contains reversing magnetic fields of 
sufficiently small scale, dissipation can take place directly
through reconnection (Drenkhahn 2002; DS02; 
Thompson 2006). Here, I focus on the reconnection model, which
makes clear prediction for the energy dissipation as function
of radius; essential for the radiative transfer calculations presented here. 
Though the results presented here
are directly applicable to the DS02 model, qualitatively similar 
results are expected from other gradual, magnetic  dissipation models.       

\subsection{The reconnection model}

An important physical quantity in the reconnection model is the ratio 
$\sigma_0$ of the Poynting flux to kinetic energy flux at the Alfv\'en 
radius $r_0$. This quantity parameterizes the baryon loading of the flow 
$\eta$ and determines the terminal bulk Lorentz factor of the flow 
$\gamma_\infty\sim \eta \simeq \sigma_0^{3/2}$. 
The flow must start Poynting-flux dominated with $\sigma_0\simmore 
30$ for it to be accelerated to ultrarelativistic speeds with $\gamma_\infty 
\simmore 100$ that are relevant for GRB flows.

In the reconnection model, the magnetic field in the flow changes polarity on 
a scale $\lambda$. If the magnetic field anchored in the rotating central 
engine is nonaxisymmetric, this scale is (in the central
engine frame) of the order of the light cylinder $r_{\rm l}$: $\lambda\simeq 
2\pi c/\Omega$, where $\Omega$ is the angular 
frequency of the rotator. This is as in the oblique rotator model 
for pulsar winds (Coroniti 1990). This model has been 
further developed to include special relativistic
effects and different reconnection prescriptions (Lyubarsky \& Kirk 2001; DS02;
Kirk \& Skj{\ae}raasen 2003). 
The rate of magnetic reconnection DS02 model is parameterized through the 
velocity $v_{\rm r}$ with which magnetic fields of opposite direction
merge. The $v_{\rm r}$ is assumed to scale with the Alfv\'en speed, $v_{\rm
  A}$, i.e. $v_{\rm r}=\varepsilon v_{\rm A}$.  A nominal value used for $\epsilon$ 
is 0.1 (see Lyubarsky 2005). For the flows with $\sigma_0\gg 1$  that are of interest 
here, the energy density of the magnetic field is larger than the rest mass 
energy density, hence $v_{\rm A}\approx c$, and the reconnection takes 
place with subrelativistic speeds.

\subsubsection{Properties of the flow}

In the reconnection model, magnetic dissipation takes place 
all the way from the initial radius $r_0$ till the saturation radius $r_s$. 
Part of the dissipated energy (approximately half) is directly used to accelerate the flow. 
The  acceleration of the flow is gradual following the 
$\gamma\sim r^{1/3}$ scaling as function of radius in the regime $r_0\ll r\ll
r_{\rm s}$. To first order approximation, no further acceleration takes place 
beyond the saturation radius.  Summarizing the results derived in Drenkhahn
(2002), the bulk Lorentz factor of the flow is approximately given by
\begin{eqnarray}
\gamma&=&\gamma_\infty\left(\frac{r}{r_{\rm s}}\right)^{1/3}=148~r_{11}^{1/3}(\varepsilon \Omega)_3^{1/3}
\sigma_{0,2}^{1/2}, \quad\rm{for}\quad r<r_{\rm s}, \nonumber\\
&&\label{gamma}\\
\gamma&=&\gamma_\infty=\sigma_0^{3/2}, \quad\rm{for}\quad r\ge r_{\rm s}, \nonumber
\end{eqnarray}
while the saturation radius is
\be
r_{\rm s}=\frac{\pi c \gamma_\infty^2}{3 \varepsilon\Omega}; \quad{\rm or}\quad
r_{{\rm s},11}=310~\frac{\sigma_{0,2}^3}{(\varepsilon \Omega)_3}.
\label{rsatur}
\ee
The notation $A=10^xA_x$ is used; the `reference values' of the
model parameters are $\sigma_0=100$, $\varepsilon=0.1$, $\Omega=10^4$
rad$\cdot$s$^{-1}$. The product of $\varepsilon$ and $\Omega$ parameterizes the
reconnection rate.  The physical quantities of the flow 
depend on this product. 

In the steady, spherical flow under consideration the comoving number density can be written as
\be n'=\frac{L}{r^2\sigma_0^{3/2}\gamma m_{\rm p}c^3}, \label{ncom}\ee
where $L$ is the luminosity per steradian of the GRB flow. 
The reference value used is $L=10^{52}$
erg$\cdot$s$^{-1}\cdot$sterad$^{-1}$.
(In this form the expression can be compared with the fireball model, where the baryon loading
parameter $\eta$ replaces the factor $\sigma_0^{3/2}$).

The expression (\ref{gamma}) is deviating from the
exact numerical solution presented in Drenkhahn (2002) at $r\simmore r_s$. 
The reason is that the dissipation does not
stop abruptly at $r_{\rm s}$ but there is modest energy
release at a slower rate at larger radii.
This leads to modest acceleration of the flow at $r\simmore r_{\rm s}$. 
In the following, we ignore these deviations and use the expressions (1)
and (3) for the bulk Lorentz factor and density of the flow respectively.  
This simplification facilitates the radiative transfer study in the flow. 

On the other hand, quantities of the flow such as  magnetic field strength and the 
energy dissipation rate as functions of radius need to be followed in more detail
around $r_{\rm s}$. Though, not important for the global energetics, the remaining dissipation at
the radii $r\simmore r_{\rm s}$ results in synchrotron emission that dominates the 
observed radiation in soft bands (such as optical and near ultra violet). 
This emission is mainly a result of the large emitting surface at these outer
parts of the flow. I take into account the residual dissipation, so that
to correctly describe the soft emission. In the calculations that follow I calculate
the magnetic field strength and dissipation rate by solving the 
relevant differential equations that describe the full 1D relativistic MHD
problem as presented in Drenkhahn (2002; summarized by Eq.~(38) in that
paper). Still, for the purpose of estimates, I give analytic expressions 
for these quantities that are accurate below the saturation radius.  

The comoving magnetic field strength $B'$ below the saturation radius
is given by setting $L\simeq L_{\rm p}=r^2B'^2\gamma^2c/4\pi$ and solving
for $B'$:
\be B'=\Big(\frac{4\pi L}{cr^2\gamma^2}\Big)^{1/2} \quad\rm{for}\quad r<r_{\rm
  s}.\label{B}
\ee
The rate of energy density release in a comoving
frame can be found by the following considerations. The time scale
over which the magnetic field decays is that of advection of 
magnetic field of opposite polarity to the reconnection area.
The reconnection speed is $v_{r}=\varepsilon v_{\rm A} \simeq \varepsilon c$,
while the magnetic field changes polarity over a comoving length scale
$\lambda'=2\pi \gamma c/\Omega$. The decay timescale for the magnetic field, therefore, is
\be
t_{\rm {dec}}=\frac{\lambda'}{v_r}=\frac{2\pi\gamma}{\varepsilon\Omega}.
\ee  
Using the last expression and Eqs. (\ref{gamma}) and (\ref{B}), 
the rate of dissipation of magnetic energy density in the comoving frame is 
\be
P_{\rm{diss}}=\frac{(B')^2/8\pi}{t_{\rm {dec}}/2}=\frac{\varepsilon\Omega L}{2\pi c
  r^2 \gamma^3}  \quad\rm{for}\quad r<r_{\rm  s}.
\label{Pdiss}
\ee  
The bulk Lorentz factor $\gamma$, the density $n'$, the magnetic field strength
$B'$ and rate of dissipation of magnetic energy density $P_{\rm{diss}}$ of the flow
as functions of radius are the quantities needed for the  study of 
the resulting emission.

\section{Photospheric and Thomson thin emission}

If all the energy is dissipated deeply into the GRB flow (i.e. at large
optical depths), adiabatic expansion converts most of this energy into
kinetic at the expense of radiation.  
Gradual dissipation heats the flow continuously and maintains a substantial 
fraction of the energy in the form of radiation.
This radiation is released at the photosphere of the flow.
If  dissipation takes place further out in the flow it
can result in additional emission coming from the Thomson thin region.
The total flux received by the observer is the integrated emission
from the different parts of the flow where dissipation takes place. 

In the case of magnetic dissipation (as well as for other dissipative
mechanisms), the fate of the released energy  
is rather uncertain. An interesting possibility
is that dissipation leads to MHD turbulence where particle acceleration can
take place by scattering of photons by Alfv\'en waves (Thompson 1994). 
On the other hand, the magnetic energy can directly 
be dissipated to the particles in the flow, most likely to the electrons due to their 
higher mobility. Following G06, GS07 we assume that a fraction $f_{\rm e}$ of
order of unity of the dissipated energy heats up the electrons. For the results
presented here we set $f_{\rm e}=0.5$.

The resulting emission does not depend only on the amounts of
energy released but also on the distribution of the emitting particles.  
I assume that the electron distribution is thermal
{\it throughout} the region where dissipation takes place. 
As I discuss in more detail in Sect. 4.2.1., the thermalization
of the electrons is result of Coulomb collisions 
in the inner parts of the flow and of exchange of synchrotron photons 
at the outer parts.      
I first summarize the results of G06, GS07 on the photospheric emission
from the reconnection model and then turn to the study of the Thomson thin
emission.

\subsection{Photospheric emission}

In addition to the saturation radius $r_{\rm s}$,
another characteristic radius of the flow is the Thomson photosphere. 
The Thomson optical depth as a function of radius is
$\tau \sim n'\sigma_{\rm T}r/\gamma$. It can be expressed
in terms of the parameters of the flow (e.g. G06):
\be
\tau=\frac{20}{r_{11}^{5/3}}\frac{L_{52}}{(\varepsilon \Omega)_3^{2/3}
\sigma_{0,2}^{5/2}}.
\label{tau}
\ee
As expected, at small radii to optical depth is large and vice-versa.
The radius of the Thomson thick-thin transition is found by 
setting $\tau=1$ in Eq.~(7) and solving for $r_{\rm ph}$:
\be
r_{\rm{ph,11}}=6\frac{L_{52}^{3/5}}{(\varepsilon \Omega)_3^{2/5}
\sigma_{0,2}^{3/2}}.
\label{rphot}
\ee
In deriving these expressions, we have assumed that  $r_{\rm{ph}}<r_s$.
A similar calculation gives the radius of the photosphere in the
 $r_{\rm{ph}}>r_s$ case.

One can check that for a large parameter space relevant 
for GRB flows, $r_{\rm {ph}}<r_{\rm s}$ which means that
dissipation proceeds throughout the photospheric region.
In terms of the physical properties of the flow, there is a critical value 
of the magnetization  $\sigma_{\rm {0,cr}}$, for which 
$r_{\rm ph}=r_{\rm s}$. For $\sigma_0>\sigma_{\rm {0,cr}}$, 
$r_{\rm ph}<r_{\rm s}$. Using Eqs.~(\ref{rphot}) and (\ref{rsatur}) one finds
\be
\sigma_{\rm {0,cr}}=42 \big(L_{52}(\varepsilon \Omega)_3\big)^{2/15}.
\label{sigmacr}
\ee
The critical baryon loading depends weakly on the parameters of the flow:
$\eta_{\rm cr}=\sigma_{\rm {0,cr}}^{3/2}=270\big(L_{52}(\varepsilon \Omega)_3\big)^{1/5}$.

For $\sigma_0\ll \sigma_{\rm {0,cr}}$ dissipation ceases
deep in the flow (at high optical depths). For  $\sigma_0\gg \sigma_{\rm {0,cr}}$,
dissipation takes place in both Thomson thick and thin
conditions with most of the energy released in the outer
parts of the flow.
The implications of such energy release to the properties
of the flow and the resulting radiation have been studied in
G06 and GS07. Those studies focused on the Thomson thick part
of the flow and the photospheric region out to $\tau \sim 0.1$.
Observational effects from Thomson thin dissipation were not
considered in detail and are the topic of this paper.

The main results of the G06, GS07 papers are the following.
Particles and radiation are found to be in thermal equilibrium deep in the flow.
There, the comoving temperature $T_{\rm{th}}$ of the flow is calculated, 
under the assumption of complete thermalization, by integrating the energy 
released at different radii in the flow and taking into account adiabatic 
cooling. Due to the dominance of scattering, the details of radiative transfer 
become important already at fairly large optical depth in the flow. 
Equilibrium between radiation and matter holds only at Thomson depths 
greater than about 50. At smaller optical depths  the electron distribution 
stays thermalized, but is out of equilibrium with the photon field. More
discussion on the processes that lead to thermalization of the electron
distribution is presented in the next section. Compton scattering of the
photons is treated in detail in this region with Monte Carlo Comptonization 
simulations (G06; GS07). Energy dissipation at moderate and low optical
depths is shown to lead to emission that has a highly non-thermal appearance. 

For $\sigma_0\simmore \sigma_{\rm {0,cr}}$, the flow develops a hot 
photospheric region.
The electron temperature at moderate optical depths can be estimated
analytically by balancing the heating [Eq.~(\ref{Pdiss})] with the inverse 
Compton cooling rate $P_{\rm Comp}=4k_{B}T_{\rm e}n'c\sigma_{\rm T}u_{\rm
  r}/m_{\rm e}c^2$, where $u_{\rm r}$ is the energy density of radiation. 
The electron temperature can be expressed as a function of 
optical depth in the flow (the detailed numerical results verify these
 estimates; see G06):
\be
T_{\rm e}\simeq \frac{40f_{\rm e}}{\tau} \quad{\rm keV,\quad for} \qquad 0.1\simless \tau
\simless 50. 
\label{Te}
\ee
For $\tau\simmore 50$, the electron temperature equals that of the photon field.
For $\tau\simless 0.1$, the electrons become mildly relativistic and one
has to consider relativistic corrections to the Compton cooling to derive the appropriate
expression for the temperature. At these larger radii (or lower $\tau$) the 
electron temperature becomes high enough that synchrotron emission 
(and associated cooling of the particles) has to be taken into account.

\subsection{Emission from the Thomson thin region}

Magnetic dissipation around the Thomson photosphere
leads to subrelativistic electrons that are more energetic with respect to the
average photon in the flow. Under these conditions
the dominant  mechanism that shapes the spectrum is inverse Compton
scattering. Synchrotron emission is negligible
since it is strongly self absorbed.   
When dissipation continues at the optically thin parts of the flow
the electron temperature becomes of the order of the electron rest mass.
At those temperatures, synchrotron self Compton emission has 
important effect on the emitted spectrum (see also Stern \& Poutanen 2004). 
Furthermore, SSA  affects the electron distribution in the flow.

\subsubsection{Thermalization of the electrons} 

At high optical depths, the density of the flow is high and the 
electron temperature rather low [see Eq.~(\ref{Te})]. Under these
conditions, Coulomb collisions are efficient in thermalizing the
electron distribution. This is not the case
in the Thomson thin parts of the flow (G06). On the other hand, there
is a more efficient channel for energy exchange among electrons.
It is shown in Ghisellini et al. (1998)  that electrons with energy of the
order of their rest mass can thermalize on a few synchrotron cooling 
times (as defined for thin synchrotron emission) by emitting and absorbing 
synchrotron photons. In the magnetically dominated flow under consideration, 
the thermalization of the electrons takes place on a timescale shorter than 
the heating/cooling one. One can, therefore, assume that the electrons 
are approximately thermal when calculating their emission.       

A caveat in the previous argument is connected to the underlying assumption
that the dissipated energy is distributed to a large fraction of the particles
in the flow. If the magnetic reconnection leads, for example, to deposition
of most of the energy to a small fraction of the electrons, they
can be accelerated to relativistic speeds and cool efficiently through
thin synchrotron emission. In this case, synchrotron self absorption is
not an efficient thermalization mechanism. This case has been investigated
in Giannios \& Spruit (2005). From this point on we assume that the dissipated
energy is distributed among a large fraction of the particles and hence thermalization
is achieved in the electron distribution.     

\subsubsection{Modeling of the synchrotron emission} 

The electron temperature becomes mildly relativistic at small 
optical depths $\tau\sim 0.1$ (see Eq.~(\ref{Te})). It increases further 
at larger radii resulting in
substantial synchrotron emission. The synchrotron-self-absorbed emission 
from mildly relativistic plasma has a characteristic spectrum that 
consists by a Rayleigh-Jeans
part up to the so-called turnover frequency $\nu_t$ where the 
optical depth due to synchrotron absorption of the flow becomes unity. Most of the energy
is emitted at the turnover frequency. At higher frequencies, the 
spectrum is very steep following the exponentially decaying tail of the 
synchrotron thin emission. The energy density per unit frequency of 
the synchrotron photons in a frame comoving with the flow is $u_\nu=8\pi 
\nu^2k_BT/c^3$ for $\nu\le \nu_t$, while it drops very fast for $\nu>\nu_t$. The 
energy density of the synchrotron photons is given by integrating the
last expression: $u_{\rm s}=\int {u_{\nu}\rm d\nu}\simeq 8\pi 
\nu_{\rm t}^3k_BT/3c^3$. The synchrotron luminosity per steradian at radius $r$ is:
\be 
L_{\rm s}=\frac{4r^2\gamma^2 u_{\rm s}c}{3}\simeq \frac{32\pi 
r^2\gamma^2\nu_{\rm t}^3k_BT_{\rm e}}{9c^2}.
\label{Ls}
\ee 
  
The synchrotron emission depends strongly on the turnover frequency.  
The turnover frequency can be related to the magnetic field strength $B'$, 
the electron temperature $T_{\rm e}$ and the Thomson optical depth $\tau$
of the flow. For the  mildly relativistic plasma
under consideration the calculation of the synchrotron emission and the
corresponding absorption is rather evolved. On the other hand, there are 
studies focusing on developing approximate expressions for the
synchrotron emission (and therefore absorption) at this temperature regime (e.g. Petrosian 1981,
Mahadevan \& Narayan 1996; Zdziarski et al. 1998; Wardzi{\'n}ski \& Zdziarski 2000). Here 
I use the approximate analytic expressions for $\nu_{\rm t}$ derived in 
Zdziarski et al. (1998): 
\be
\nu_{\rm t}=\frac{343}{36}\theta_{\rm e}^2\nu_{\rm
  c}\ln^3\frac{C}{\ln{\frac{C}{\ln \frac{C}{...}}}},
\label{nut}
\ee
where $\nu_{\rm c}=eB'/2\pi m_{\rm e}c$ is the cyclotron frequency and 
$\theta_{\rm e}=k_{\rm B}T_{\rm e}/m_{\rm e}c^2$. 
The quantity $C$ is defined as\footnote{Here I set the Zdziarski et al. (1998)
 correction factor $A_{\rm M}=1$.}
\be
C=\frac{3}{7\theta_{\rm e}}\Big[\frac{\pi m_{\rm e}c^2\tau e^{1/\theta_{\rm e}}}
{3\alpha_f h\nu_{\rm c}}\Big]^{2/7},
\label{c}
\ee
where $h$, $\alpha_f$ are Planck's and fine structure constants respectively. 
For typical model parameters and radii $r\sim r_{\rm s}$, the turnover
frequency is $\nu_{\rm t}\sim 10^3 \nu_{\rm c}$ which results in 
$\nu_{\rm t}^{\rm obs}=\gamma_{\infty} \nu_{\rm t}/(1+z)\simless 1$ keV
($z$ is the redshift of the burst).
The synchrotron-self-absorbed emission appears mainly in the soft X-rays
and softer bands. 

The expression (\ref{nut}) has been compared with the exact numerical results
and shown to be overestimating the turnover
frequency (and consequently the synchrotron emission) 
in plasma with $\theta_{\rm e}\simless 0.1$ (see fig.~2b in  Wardzi{\'n}ski 
\& Zdziarski 2000). On the other hand, it is quite accurate for
higher electron temperatures. Since most of the synchrotron emission
comes from regions of the flow with $\theta_{\rm e}\sim 1$, Eq.~(\ref{nut})
is accurate enough for the calculations presented here.

One can now extend the Comptonization calculation developed in G06
to include the radiative transfer in the Thomson thin region of the flow. The procedure is the following.
I make a choice for the temperature profile as function of radius of the electron temperature
in the flow. Analytical expressions such as that of Eq.~(10) provide a good
initial guess. The radiative transfer in the flow is studied using the Monte
Carlo Comptonization code described in G06. The calculation  includes 
the thermal radiation field carried with the flow which is injected 
in the inner numerical boundary at the ``equilibrium'' radius
where radiation and particles drop out of equilibrium. 
At larger radii the synchrotron emitted flux is also
included (using expressions (\ref{Ls}), (\ref{nut})). 
Both sources of photons are propagated through the medium and their
scattering by electrons is followed. The code calculates
the spectrum and the radius-dependent cooling rate of the electrons.
Cooling because of inverse Compton, synchrotron emission and adiabatic expansion
is taken into account. The adiabatic cooling becomes important at high optical
depths and at the very outer parts of the flow 
where the expansion timescale $r/\gamma c$ is shorter than the radiative cooling one.
The outer boundary of the calculations is set at large enough radius so that
it does not have an effect on the computed spectra. 
I iterate the electron temperature until the cooling rate matches 
the heating rate predicted by the model reasonably well at all radii. 
In practice, I make sure that they match within $\sim 30$\% or less everywhere. 
Considering the rather larger uncertainties in the model
coming from, for example, the assumed dynamics of the flow, this a fairly accurate
calculation.     

\section{Resulting spectra}

First, I focus on the new features that appear in the emitted spectrum
w.r.t. the G06 results because of the inclusion of the Thomson thin emission.
In the illustrative case of Fig.~1, I set $\sigma_0=70$ (which corresponds to
flow with baryon loading $\eta\simeq \sigma_0^{3/2}\sim 600$) and the rest
of the parameters to their reference values. Spectra are 
plotted in the central engine frame. The dotted line shows the
input thermal radiation at the ``equilibrium radius'' which is the inner boundary
of the computed domain. The thermal flux is advected with the flow
and constitutes a large fraction of the seed photons to be Comptonized 
further out in the flow. 

The appearance of the photon spectrum at radius 
which corresponds to optical depth $\tau=0.1$ is shown with dashed line. This radius
is the outer boundary used in most of the calculations of G06, GS07.
One can clearly see the effect of inverse Compton scattering to the spectrum.
The peak of the $E\cdot f(E)$ spectrum increases slightly and the spectrum
becomes broader. An important feature is the high-energy tail that
is result of unsaturated Comptonization at $\tau\simless 1$.
Note also a second peak at $\sim$1 keV. This is result of
synchrotron emission with the turnover frequency being
$\sim$1 keV (in the central engine frame). This component is still weak at $\tau=0.1$
and has not been included in the calculations of G06, GS07.

\begin{figure}
\resizebox{\hsize}{!}{\includegraphics[angle=270]{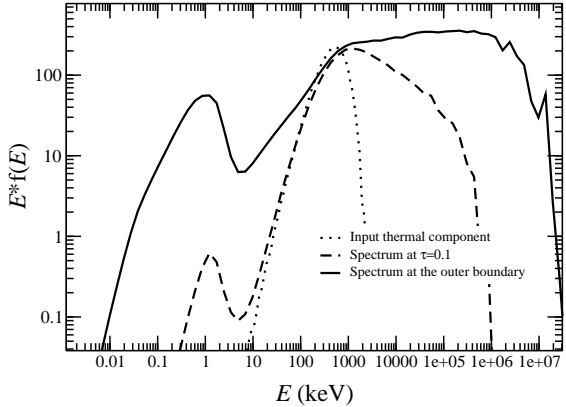}}
\caption[] {Photon spectrum at different radii (or corresponding Thomson
optical depths) in the flow. The spectrum is shown in the central engine frame.
The dotted line stands for the spectrum at the radius where radiation 
and electrons decouple. The photospheric emission is shown
with the dashed line (see also G06; GS07).
The overall spectrum (solid line) includes the emission from the
Thomson thin region of the flow. Synchrotron-self-absorbed 
emission from this region dominates the spectrum below $\sim 10$ keV.
Inverse Compton leads to flat $\gamma$-ray tail up to $\sim$ 1 GeV.
\label{fig1}}
\end{figure}

The total spectrum is shown with solid line. This includes the emission from
the whole volume of the flow where there is dissipation taking place.
The overall emission spectrum is much broader. It is characterized by a break at
$\sim 1$ MeV followed by a flat spectrum with photon-number index
$\Gamma\simeq -2$ (where d$N/$d$E\sim E^\Gamma$), close to the typically
observed one. The hard $\gamma$-ray tail is
extending up to $\sim$1 GeV which corresponds to the Lorentz boosted
temperature of the flow at its outer layers (where it reaches its maximum
values). Comparing the spectrum at $\tau=0.1$ and the total one, it is 
clear that Comptonization proceeds throughout the
Thomson thin region strengthening the hard $\gamma$-ray component.

An important  new feature is the powerful emission that appears in the soft X-rays and softer bands.
This comes from synchrotron-self-absorbed emission. SSA dominates by many
orders of magnitude the ultra violet and optical emission.
 The softer emission originates from the Thomson thin part of the flow that is 
characterized by the higher electron temperatures and larger emitting surface
(see also Stern \& Poutanen 2004). This emission is very weak in models where
dissipation takes place below or around  the Thomson photospheric region
(see, e.g., M\'esz\'aros \& Rees 2000; Pe'er et al. 2006; Ioka et al. 2007).

As a result of the synchrotron-self-Compton component, the spectrum below the MeV 
peak softens with respect to the G06 calculation where only the photospheric
component was considered. The spectrum can be well fitted with a power-law in the $30-300$ keV
energy range with photon-number index of $\Gamma\simeq -1.2$ which
very close to the one typically observed (e.g. Preece et al. 1998).  

In this example, the SSA emission spectrum hardens considerably below $\sim
30$ eV. At lower energies  the Rayleigh-Jeans limit is gradually approached.
The location of the hardening is determined by the radius where
adiabatic expansion starts to dominate the cooling of the electrons.    
In this example adiabatic expansion dominates at $r\simmore r_{\rm ad}\sim 5\cdot 10^{14}$ cm. 
Most of the radiation observed at $E\simless 30$ eV comes from this radius.
The optical and near UV emission is delayed w.r.t. the $\sim$ MeV emission by
$\delta t\sim r_{\rm ad}/\gamma_\infty^2c\sim 0.05$ s. Radiation in these
bands reaches the observer with small but maybe detectable lags.  
In the region $30\simless E\simless 1000$ eV the spectral slope depends on the radial
dependence of the turnover frequency $\nu_{\rm t}$. 

The relative strength of the synchrotron-self-Compton (SSC) component depends on the fraction of energy 
dissipated in the Thomson thin region of the flow. For $\sigma_{0}\gg
\sigma_{\rm {0,cr}}$ most of the
energy is dissipated in the Thomson thin region. Correspondingly the SSC
component is pronounced. This is evident in Fig.~2 where the
spectrum is shown for different values of $\sigma_0$ (the rest
of the parameters are kept in their reference values).     
For $\sigma_0=40$, dissipation stops close to the photosphere of the flow.
The SSA component is almost absent and the emission above the thermal
peak at $\sim 1$ MeV relatively weak. In more baryon loaded models the emission is
quasi-thermal since dissipation stops at high Thomson depths.

\begin{figure}
\resizebox{\hsize}{!}{\includegraphics[angle=270]{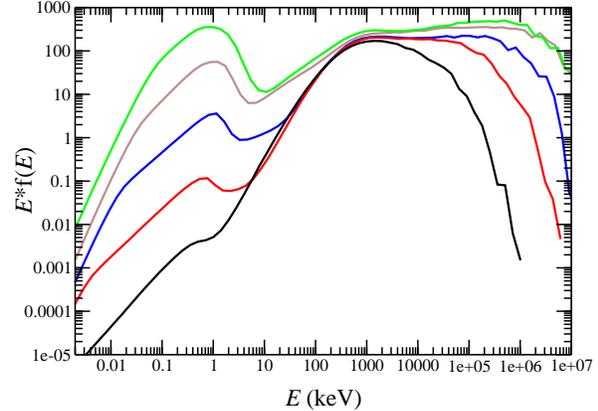}}
\caption[] {Resulting spectrum (in the central engine frame)
for different  baryon loadings of the flow. From bottom to top the curves correspond to
magnetization $\sigma_0=$40, 50, 60, 70, 100 
(or corresponding baryon loading $\eta\simeq$ 250, 350, 460, 590, 1000)
respectively. The high $\sigma_0$ flows  
are characterized broader spectra. The model predicts that
bright prompt optical and UV emission is accompanied by 
powerful $\sim$GeV emission. For bright optical emission,
the optical spectrum is expected to be hard.

\label{fig1}}
\end{figure}

With increasing $\sigma_0$ both the SSA and inverse Compton components
become relatively more powerful. The thermal peak is followed by a
flat $\gamma$-ray emission that extends up to $\sim $0.1-1 GeV.
Most of the models show a high energy cutoff in this energy range.
This cutoff corresponds to the highest energies to which photons
are upscattered. It is determined by the Lorentz boosted electron
temperature at $r\sim r_{\rm s}$.
The spectral slope below the $\sim$1 MeV break becomes softer
with increasing $\sigma_0$. The photon-number index in the
$30-300$ keV energy band varies in the range
$-1.2\simless \Gamma\simless -0.4$ in agreement to that typically
observed (e.g. Preece et al. 1998). 
The high $\sigma_0$  models have powerful optical and near ultra violet 
emission. The flux $f(E)$ that is emitted in these bands is
similar to the X-rays one.  The optical spectrum is
hard with photon number index $0\simless \Gamma\simless +1$.

Varying the baryon loading of the flow has moderate effect in the emission
in the {\it BATSE} energy range but profound implications in other bands.  
The model predicts that flows with low baryon loading (i.e. high  $\sigma_0$)
have powerful optical, UV and GeV emission. 
More on the comparison of the model with observations is presented in the next section. 

\subsection{Comparison with observations}

The prompt GRB emission has been typically observed in the hard X-rays 
up to $\sim 1$ MeV $\gamma$-rays. The spectrum in this energy range 
shows a characteristic sub-MeV break followed by a flat power-law $\gamma$-ray 
tail (e.g. Band et al. 1993). Below the break the spectrum has typical photon number index of 
$\Gamma\sim -1$ although much harder spectra have also been observed\footnote
{These hard spectra cannot be explained by the thin synchrotron model
(e.g. Ghirlanda, Celotti \& Ghisellini 2003).}. 
The observed sub-MeV break and the spectral slopes above and below the break
 are naturally explained by the gradual dissipation
model discussed here. Furthermore, the
model makes specific predictions on the prompt emission from the 
optical to GeV; bands that are currently (or will soon be) accessible to observations.
The model predicts that the flat $\gamma$-ray tail extends up to a cutoff that
typically appears at $\sim 0.1-1$ GeV. It also predicts the prompt optical and UV
emission. For low baryon loading, the emission in these bands  is
powerful with energy flux $f(E)$ similar to that of X-rays. The optical emission is 
characterized by a hard spectrum. Optically bright bursts have powerful
GeV emission and softer spectra below the $\sim$1 MeV break. 

In recent years, several observations in softer bands have been made
simultaneously with the prompt GRB emission.
The {\it Swift} satellite has observed the prompt emission in 
the X-rays (Hill et al. 2006; Romano et al. 2006; Page et
al. 2007) and ultra violet (Page et al. 2007) and robotic telescopes 
in the optical and infra red (e.g. Akerlof 1999; Vestrad 2005; 
Blake 2005; Bo\"er et al. 2006; Vestrand et al. 2006; Klotz et al. 2006). 
Furthermore, {\it GLAST} is expected to probe the emission from 
GRBs up to $\sim 100$ GeV.

\begin{figure}
\resizebox{\hsize}{!}{\includegraphics[angle=270]{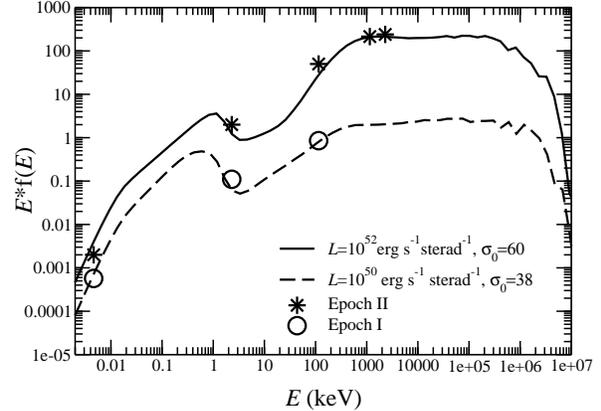}}
\caption[] {Applying the model to multi frequency observations 
of the prompt emission of GRB 061121 (see Fig. 11 in Page et al. 2007). 
The circles stand for the observations of epoch I (just before the main pulse of the burst) 
and the stars for those of epoch II (during the pulse). 
Observations are blue-shifted by 1+$z$ to the burst rest frame 
($z=1.131$ for GRB061121). The solid and dashed curves show spectra 
for two different sets of the parameters of the flow that illustrate 
that the model can account for the broad-band prompt spectra.
 
\label{fig1}}
\end{figure}

Here, I compare the model to the very well sampled prompt 
emission of GRB 061121. This burst has been observed from optical 
to $\sim 1$ MeV (Page et al. 2007). The prompt emission has been followed with
XRT and UVOT on board to {\it Swift} and in the optical with {\it ROTSE} simultaneously 
to $\gamma$-ray observations with BAT and {\it Konus-Wind}. There are two time resolved
spectra just before and during the main pulse of the prompt emission 
that appears $\sim 75$ sec after the onset of the burst. The pulse is clear
in the lightcurves in all observed energy bands. The correlation between
the different bands indicates that the optical, UV and X-ray and $\gamma$-ray components 
have common origin (i.e. they are connected to the prompt GRB).  
This is unlike cases where the optical lightcurves are not tracing
the $\gamma$-rays (e.g. Akerlof 1999; Bo\"er et al. 2006; Klotz et al. 2006 ) 
suggestive of a different physical origin w.r.t. that of the prompt emission.

In Fig.~3, the data shown with circles refer to the pre-spike emission (epoch I in the 
Page et al. 2007 terminology) and the stars to the peak observed luminosity 
of the burst (epoch II). The data span approximately 6 orders of magnitude in frequency
from the optical to $\sim 1$ MeV. Overplotted are the spectral predictions
of the model for two different sets of parameters. The low luminosity model
has $L=10^{50}$ erg$\cdot$s$^{-1}\cdot$sterad$^{-1}$ and $\sigma_0=38$ and the high luminosity
one $L=10^{52}$ erg$\cdot$s$^{-1}\cdot$sterad$^{-1}$ and $\sigma_0=60$. The two models (not meant to
be detailed fits) are reproducing the observations quite closely.

Note that for a given observed luminosity, the baryon loading is essentially
the only free parameter of the model. This can be constrained by the 
ratio of the $\sim 1$MeV-to-optical flux. Additional 
constraints can come from observations of the prompt emission in 
harder bands. In this respect {\it GLAST} observations  in the $\sim$GeV 
range are going to be of particular importance.

The high luminosity model (that describes the epoch II
observations) is characterized by higher $\sigma_0$ with respect to the lower 
luminosity one. This is in qualitative agreement with the baryon loading-luminosity
correlation during the evolution of the burst needed to explain
observed energy-dependent properties of the GRB pulses in the context of
the reconnection model (for details see sect. 4 in GS07). 
However since GS07 do not consider the Thomson thin emission
in the calculations, the quantitative results of section 
4 in GS07 have to be revisited.

\section{Discussion/conclusions}
\label{conclude}

The GRB emission may be result of internal collisions in 
a variable flow (Rees \& M\'esz\'aros 1994). 
In the internal shock model energy is dissipated at the
location of the shell collision. Particles are accelerated at the
shock front to ultrarelativistic speeds and non-thermal distributions.
Magnetic fields are assumed to be amplified because of
plasma instabilities. The emitted spectrum depends on the
distribution of the emitting particles and the strength of the
shock amplified magnetic field, both of which are not understood 
from first principles. The emitted spectrum also depends on the 
properties of the flow (such as optical depth) at the radius of 
the collision (e.g. Pe'er et al. 2006).

As an alternative to internal shocks,  magnetic dissipation
in a strongly magnetized flow can power the GRB (Thompson 1994).
Magnetic dissipation may lead to gradual
release of energy over a wide range of radii (e.g. Drenkhahn 2002; DS02).
It typically proceeds in both Thomson thick and thin regions of the flow. 
The released energy can be distributed to a large fraction of the particles 
of the flow leading to the slow heating scenario for the GRB emission 
(Ghisellini \& Celotti 1999). The emitting particles (i.e. electrons)
are heated up to mildly relativistic speeds. Because of the strong magnetic 
fields exchange of synchrotron photons provides an efficient mechanism for 
the thermalization of the electron distribution (Ghisellini et al. 1998).
Since the emitting particles are thermal the resulting emission does not
depend sensitively on poorly understood physics of particle acceleration in magnetic
reconnection. The model is defined by just the luminosity of the flow,
its baryon loading and a reconnection-rate parameter and makes
direct and stable predictions for the electromagnetic spectrum.

In previous works (G06 and GS07), we calculated the photospheric emission 
expected from the reconnection model.
The radiative transfer study was made with Monte Carlo simulations.
Those calculations have shown that the  flow is characterized by
powerful photospheric emission  with most of the energy
appearing in the hard X-rays and $\sim1$ MeV $\gamma-$rays.
This emission is the result of photons, produced deep into the flow, that 
are inverse Compton scattered by sub-relativistic electrons at 
Thomson optical depths of order of unity.

In the reconnection model, for low enough baryon-loading of the
flow, there are substantial amounts of energy dissipated in its outer, 
Thomson thin, parts. Here I have extended the G06 calculations to include the Thomson 
thin emission. Energy release at large radii leads to mildly relativistic
electrons that cool down though emitting synchrotron radiation and inverse
Compton scattering soft photons. 
SSA emission dominates the observed radiation in the
soft X-ray and softer bands.  This soft (e.g. optical) emission is very 
weak in models where dissipation is limited below or around the Thomson 
photosphere (see, e.g., M\'esz\'aros \& Rees 2000; Pe'er et al. 2006; Ioka et al. 2007).
Inverse Compton in the Thomson thin region leads to a flat high-energy
spectrum that extends up to GeV energies.

The resulting spectra from the radiative transfer calculations naturally 
explain the observed sub-MeV break of the GRB emission and the spectral slopes
above and below the break (Band et al. 1993). Furthermore, the model makes rather robust predictions
for the emission in other energy bands. The flat $\gamma$-ray spectrum is
expected to show a cutoff in the $\sim 0.1-1$ GeV energy range that 
should be observable with {\it GLAST}. The optical and ultra violet 
emission can be powerful and the optical spectrum hard with 
photon number index $0\simless \Gamma\simless 1$.
Bright prompt optical emission is predicted to be
accompanied by powerful $\sim$ GeV emission and rather soft
spectra below the sub-MeV break. 
Comparison with multi-frequency observations of the
the prompt emission from GRB 061121 that span from the optical to
the $\sim$ MeV range (Page et al. 2007) supports the model.       

\begin{acknowledgements}
I thank Henk Spruit for many valuable suggestions and discussions. 
\end{acknowledgements}


\begin{thebibliography}{}

\bibitem{} Akerlof, C., et al.\ 1999, \nat, 398, 400 
\bibitem{} Band, D., Matteson, J., Ford, L., et al. 1993, ApJ, 413, 281
\bibitem{} Blake, C.~H., et al.\ 2005, \nat, 435, 181
\bibitem{} Bo{\"e}r, M., Atteia, J.~L., Damerdji, Y., Gendre, B., Klotz, 
A., \& Stratta, G.\ 2006, \apjl, 638, L71 
\bibitem{} Coroniti, F. V. 1990, ApJ, 349, 538 
\bibitem{} Daigne, F., \& Mochkovitch, R.\ 1998, \mnras, 296, 275 
\bibitem{} Drenkhahn, G.\ 2002, \aap, 387, 714
\bibitem{} Drenkhahn, G., \& Spruit, H. C.\ 2002, \aap, 391, 1141 (DS02)
\bibitem{} Ghirlanda, G., Celotti, A., \& Ghisellini, G.\ 2003, \aap, 406, 879
\bibitem{} Ghisellini, G., Haardt, F., \& Svensson, R.\ 1998, \mnras, 297, 348 
\bibitem{} Ghisellini, G., \& Celotti, A. 1999, A\&AS, 138, 527
\bibitem{} Giannios, D.\ 2006, \aap, 457, 763 (G06)
\bibitem{} Giannios, D., \& Spruit, H. C.\ 2005, \aap, 430, 1
\bibitem{} Giannios, D., \& Spruit, H. C.\ 2006, \aap, 450, 887
\bibitem{} Giannios, D., \& Spruit, H.~C.\ 2007, \aap, 469, 1 (GS07)
\bibitem{} Ioka, K., Murase, K., Toma, K., Nagataki, S., \& Nakamura, T.\ 2007, \apjl, 670, L77    
\bibitem{} Katz, J.~I.\ 1994, \apjl, 432, L107 
\bibitem{} Kirk, J.~G., \& Skj{\ae}raasen, O.\ 2003, \apj, 591, 366
\bibitem{} Klotz, A., Gendre, B., Stratta, G., Atteia, J.~L., Bo{\"e}r, M., 
Malacrino, F., Damerdji, Y., \& Behrend, R.\ 2006, \aap, 451, L39 
\bibitem{} Kobayashi, S., Piran, T., \& Sari, R.\ 1997, \apj, 490, 92  
\bibitem{} Lyubarsky, Y. E. 2005, MNRAS, 358, 113
\bibitem{} Lyubarsky, Y., \& Kirk, J.~G.\ 2001, \apj, 547, 437 
\bibitem{} Lyutikov, M., \& Blandford, R.\ 2003, ArXiv e-prints, arXiv:astro-ph/0312347 
\bibitem{} Mahadevan, R., Narayan, R., \& Yi, I.\ 1996, \apj, 465, 327
\bibitem{} M\'esz\'aros, P., \& Rees, M. J.\ 2000, \apj, 530, 292
\bibitem{} Mimica, P., Aloy, M.~A., M{\"u}ller, E., \& Brinkmann, W.\ 2005,
  \aap, 441, 103 
\bibitem{} Nakar, E., \& Piran, T.\ 2002, \apjl, 572, L139
\bibitem{} Page, K.~L., et al.\ 2007, \apj, 663, 1125
\bibitem{} Paczynski, B., \& Xu, G.\ 1994, \apj, 427, 708 
\bibitem{} Pe'er, A., M{\'e}sz{\'a}ros, P., \& Rees, M.~J.\ 2006, \apj, 642, 995
\bibitem{} Petrosian, V.\ 1981, \apj, 251, 727 
\bibitem{} Preece, R. D., Briggs, M. S., Mallozzi, R. S, et al. 1998, \apj,
  506, L23
\bibitem{} Rees, M. J., \& M\'esz\'aros, P. 1994, ApJ, 430, L93
\bibitem{} Spruit, H. C., Daigne, F., \& Drenkhahn, G. 2001, A\&A, 369, 694
\bibitem{} Stern, B.\ 1999, in ASP Conf. Ser. 161, High Energy Processes in Accreting Black Holes,
  ed. J. Poutanen, \& R. Svensson (ASP San Francisco), 277 
\bibitem{} Stern, B.~E., \& Poutanen, J.\ 2004, \mnras, 352, L35 
\bibitem{} Tavani, M.\ 1996, \apj, 466, 768 
\bibitem{} Thompson, C.\ 1994, \mnras, 270, 480
\bibitem{} Thompson, C.\ 2006, \apj, 651, 333
\bibitem{} Vestrand, W.~T., et al.\ 2005, \nat, 435, 178
\bibitem{} Vestrand, W.~T., et al.\ 2006, \nat, 442, 172 
\bibitem{} Wardzi{\'n}ski, G., \& Zdziarski, A.~A.\ 2000, \mnras, 314, 183 
\bibitem{} Zdziarski, A.~A., Poutanen, J., Mikolajewska, J., Gierlinski, M., 
Ebisawa, K., \& Johnson, W.~N.\ 1998, \mnras, 301, 435 

\end{thebibliography}
\end{document}